\documentclass[technote, 10pt]{IEEEtran}
\IEEEoverridecommandlockouts
\usepackage{cite}
\usepackage{amsmath,amssymb,amsfonts}
\usepackage[mathscr]{euscript}
\usepackage{algorithmic}
\usepackage{graphicx}
\usepackage{textcomp}
\usepackage{xcolor}
\usepackage{booktabs}
\usepackage{subcaption}
\usepackage{xfrac}

\usepackage{nicematrix}

\def\BibTeX{{\rm B\kern-.05em{\sc i\kern-.025em b}\kern-.08em
    T\kern-.1667em\lower.7ex\hbox{E}\kern-.125emX}}
\begin{document}

\title{Instantaneous directional channel measurements at 14 GHz and 160 GHz via a virtual circular array
\thanks{The authors acknowledge the financial support by the Federal Ministry of Education and Research of Germany in the program of ``Souverän. Digital. Vernetzt.'' Joint project 6G-RIC, project identification number: 16KISK020K.}
}

\author{
\IEEEauthorblockN{Wilhelm Keusgen}
\IEEEauthorblockA{\textit{Technische Universität Berlin} \\
Berlin, Germany \\
ORCID 0000-0002-2335-8270\\}
\and
\IEEEauthorblockN{Taro Eichler}
\IEEEauthorblockA{\textit{Rohde \& Schwarz GmbH \& Co. KG} \\
Munich, Germany\\
taro.eichler@rohde-schwarz.com}
}

\maketitle

\begin{abstract} In this paper a novel frequency-scalable rotary platform design is introduced which allows for flexible directional channel measurements using different types of antennas, and which can also be used with frequency extenders for measurements up to the THz region. The measurement platform has been applied to measure the channel properties including the direction of arrival at the FR3 frequency 14 GHz and in the D-band at 160 GHz in a large hall indoor environment with LOS distances up to 40 m. The results show very good agreement of strong path components for both frequencies as well as interesting dependencies of delay spread, angular spread, and Ricean K-factor on distance and frequency and can be used to parameterize a path loss model.
\end{abstract}

\begin{IEEEkeywords}
6G, sub-THz, FR3, 14 GHz, D-Band, 160 GHz, channel measurements, direction of arrival, virtual circular array, rms delay spread, rms angular spread, Ricean K-factor, multipath components, pathloss, CI model.
\end{IEEEkeywords}


\section{Introduction}
As one of the potential frequency pillars of the first commercial 6G networks in 2030, the ITU World Radio Conference 2023 (WRC-23) has identified new frequency bands as 6G study items for the WRC-27 in the FR3 spectrum (7.125 - 24 GHz), in particular the frequency bands 7.125-8.4 GHz, 14.8–15.35 GHz and 12.7-13.25 GHz. 3GPP has already initiated a study item for FR3 channel modeling also in the context of sensing applications. While sub-THz frequencies for communication within 6G networks will potentially be introduced at a later stage, sensing applications could represent an attractive use case leveraging the higher frequencies and bandwidths to obtain a higher resolution. Furthermore, sub-THz frequencies for communication could become essential to realize the full potential of the metaverse, extended reality (XR) and enable professional outdoor use cases, providing immersive and interactive experiences critical for many industries. 6G channel modeling activities are also currently being pursued in the ETSI industry specification group ISG THz. For sensing applications and because 3D beamforming is often applied in communication at these frequencies, the directional channel information is indispensable. Here we present a frequency-scalable rotary platform which allows to obtain directional information with a very high angular resolution in an efficient and fast manner by using an improved  virtual array technique. This approach overcomes most of the disadvantages of using stepwise rotating high gain antennas (see e.g. \cite{haneda}), which are extremely low measurement speed, missing phase coherence between the different paths, as well as a complicated de-embedding of antenna patterns, although the measurement sensitivity is very good.  Another known approach is the use of switched directive antennas \cite{sun}, which have a very fast measurement time and high sensitivity, but still no coherence between different directions, the problem of antenna de-embedding and a poor angular resolution. 

In the following, we describe the improved measurement technique and demonstrate its feasibilty by comparative measurements and investigations of the directional indoor radio channel in the FR3 band at 14 GHz and the sub-THz D-band at 160 GHz.





%

\section{Measurement Approach}
We propose the use of a virtual uniform circular array antenna (VUCA) \cite{nguyen} -- implemented by an antenna moving on a circumference -- in combination with wideband time-domain channel sounding. In contrast to the original idea, which requires omni-directional antennas, the concept was extended to the use of antennas with moderate gains, like open waveguide antennas. This has multiple advantages like additional antenna gain and thus higher sensitivity as well as better availability of suitable antennas especially for vertical polarization. This approach allows in combination with a novel rotary platform, the instantaneous directional measurement of radio channels up to the THz regime and with two polarizations: Since the used beamspace multiple signal classification (BS MUSIC) \cite{mathews} works mainly on phase information and gives reasonable results even for semicircular arrays, the direction of arrival (DoA) information could be retrieved even from measurements with directional antennas, whereas the magnitude information is obtained from the respective maximum of the input delay spread function (IDSF, i.e the temporal set of channel impulse responses: CIRs), under the assumption that at each delay sample there is a maximum of one significant path. This condition might be fulfilled for broadband measurements and might be supported by an suitable interpolation in the delay domain. Due to the impact of oversampling and the necessary frequency window, which decreases the delay resolution, and also due to small changes in delay for the different antenna positions on the circumference, one significant path is usually spread over multiple adjacent delay samples. This issue can be overcome with a clustering of nearby paths in the angular-delay domain and taking the local maximum of each cluster as true value. We implemented this approach just for the azimuth angle of arrival ($\varphi$) and simplified the clustering by rounding the azimuth values to a defined grid and leaving the (oversampled) delay grid unchanged. Finally, we were able to estimate a discrete set of set of $L$ paths $\{p_l(\tau_l,\varphi_l)\}$ with powers $P_l = |p_l|^2$, delays $\tau_l$, and azimuth angles $\varphi_l$, having sub-sample delay resolution and an angular resolution defined by the clustering grid. Since the estimation still includes the antenna gain of the rotating antenna, it has to be compensated for.

One concern of the VUCA approach is the Doppler spread introduced due to the continuous movement of the antenna which might impair the estimation result. As it was shown in \cite{wittig} the approximated linear phase shift during the measurement (sampling) of one virtual antenna element, is transformed into a delay shift of the CIRs, when using Frank-Zadoff-Chu sequences with root parameter $\lambda = 1$. With the minimum required number $K_{min}$ of virtual antennas
\begin{equation}
    \label{eq:k_min}
    K_\mathrm{min} = \left\lceil\dfrac{4\,\pi\,R_A}{\lambda_0}\right\rceil
\end{equation}
for a given wavelength $\lambda_0$ and $R_A$ being the radius of the VUCA, it can be shown that the maximum expected shift in delay is $\tfrac{1}{2} T_s = \tfrac{1}{2 B}$, where $T_s$ is the sampling period and $B$ is the bandwidth. Since this effect adds up to the above mentioned spread of delays, it can also be handled by the proposed approach.

\section{Parameter Estimation}
\label{sec:parameter}
Based on the estimated discrete set of paths, the following channel parameters can be determined.
\subsection{Power and path loss}
Assuming a calibrated measurement, where 0 dB channel gain corresponds to a back-to-back connection of transmitter (Tx) and receiver (Rx) antenna ports, the overall channel power $\overline{P}$ and the path loss (PL) including antennas (radio channel) is defined to:
\begin{equation}\label{eq:P_total}
\overline{P} = \sum_{l = 1}^L P_l(\tau_l,\varphi_l) := \frac{1}{\mathrm{PL}},
\end{equation}
with $\mathrm{PL}_{|\mathrm{dB}} = -10 \,\mathrm{log}_{10} (\overline{P})$. For a line-of-sight (LoS) situation, the propagation path loss $\widehat{\mathrm{PL}} = \tfrac{1}{G_{Tx}G_{Rx}}\mathrm{PL}$ of the propagation channel can be modeled depending on distance $d$ with the close-in (CI) model according to
\begin{equation}
\label{eq:CI model}
    \widehat{\mathrm{PL}}_{|\mathrm{dB}}(d)  = 20\,\mathrm{log}_{10}\left(\tfrac{\lambda_0}{4 \pi \cdot 1 \mathrm{m}}\right)+ 10\,n\,\mathrm{log}_{10}(\sfrac{d}{1 \mathrm{m}})\text{,}
\end{equation}
with $G_{Tx}$ and $G_{Rx}$ being the antenna gains at Tx and Rx.
\subsection{Number of Paths and Ricean K-Factor}
The absolute number of paths is given by the cardinality $L$ of the estimated set $L$ paths $\{p_l(\tau_l,\varphi_l)\}$. When assuming a LoS channel with more than one path, the number of multipath components (MPCs) equals to $L-1$. In LoS the first path in the set is the direct path and the Ricean K-factor \underline{K} can be defined to be
\begin{equation}
\underline{K} = \frac{\overline{P}}{\sum_{l = 2}^L \overline{P}_l(\tau_l,\varphi_l)}
\end{equation}
with $\underline{K}_{|\mathrm{dB}} = 10 \,\mathrm{log}_{10} (\underline{K})$.
\subsection{RMS Delay Spread and RMS Angular Spread}
The root mean square delay spread (RMS DS) characterizes the probability density of the channel power with respect to delays and is defined as the second central moment of $P(\tau,\varphi)$
\begin{equation}
\label{eq:second_central_moment}
    \sigma_d^2 = \frac{1}{\overline{P}} \sum_{l=1}^{L} (\tau_l - \overline\tau_l)^2\, P(\tau_l)
\end{equation}
with
\begin{equation}\label{eq:first_raw_moment}
\overline\tau = \frac{1}{P} \sum_{m'=1}^{M'} \tau_{m'}\, \overline P(\tau_{m'}) \text{,}
\end{equation}
being the first moment of $P(\tau,\varphi)$ \cite{keusgen}.

Accordingly, the power density with respect to the azimuth angle $\varphi$ can be described as root mean square angular spread (RMS AS)
\begin{equation}
\label{eq:rms_as}
\sigma_\varphi^2 = \sqrt{-2\,\mathrm{log}\left(\dfrac{1}{\overline P}\left\lvert\sum_{l=1}^L P(\varphi_l)\,\mathrm{exp}(j \varphi_l)\right\rvert\right)}
\end{equation}
taking the circular nature of $\varphi$ into account \cite{3gpp}.

\section{Measurement Setup and Scenario}
The introduced measurement approach was experimentally validated by a channel sounder based on commercial test and measurement equipment, supporting mm-wave and sub-THz bands and a custom made rotary platform. At the Tx side a vector signal generator R\&S\textregistered SMW200A was used to generate RF signals modulated with a Frank-Zadoff-Chu sequence. At the Rx side a vector signal and spectrum analyzer R\&S\textregistered FSW was applied to down-convert and sample the received signals back into the digital base-band domain. Two frequencies points - one in the microwave band (14 GHz) and one in the sub-THz band (160 GHz) - were investigated, where the frequency range of the measurement equipment was enhanced up to the D-Band (110 - 170 GHz) using active frequency extenders on both sides (R\&S\textregistered FE170ST and R\&S\textregistered FE170SR). The VUCA was implemented by a rotary platform which is equipped with different rotary joints for RF-signals, Ethernet and DC power supplies, such that active components like frequency extenders or amplifiers could be rotated together with the antenna on freely configurable radii, enabling VUCA measurements up the THz range (300 GHz and above). Here, we chose 1000 and 1440 virtual antenna elements for measurement durations of 500 ms and 720 ms, 
and a measurement bandwidth of 2 GHz. The Table \ref{tab:cs_parameters} lists the relevant parameters of the channel sounder. Further details on time domain channel sounding can be found in \cite{keusgen}.
\begin{table}[htbp]
\caption{Parameters of the channel sounder}
\label{tab:cs_parameters}
    \begin{center}
        \begin{NiceTabular}[c]{l l l}
        \CodeBefore
        \rowcolor{gray!50}{1-1}
        \rowcolors{2}{gray!15}{}
        \Body
            \toprule
			\textit{Parameter} & \textit{FR3} & \textit{sub-THz}\\
			\midrule
            carrier frequency ($f_0$)& 14 GHz & 160 GHz\\
            measurement bandwidth ($B$)& 2 GHz & 2 GHz\\
            Rx sampling rate ($f_S$)& 2.5 GHz & 2.5 GHz\\
            sequence length ($M$) & 1,000,000  & 1,000,000\\
            \# of seq., \# of virt. antennas ($K$) & 1000 & 1440\\
            sequence duration ($T_s$) & 500 µs & 500 µs\\
            measurement time ($T_m$) & 500 ms & 720 ms\\
            oversampling gain ($10 \, \mathrm{log}_{10}(\sfrac{f_S}{B})$) & 1 dB& 1 dB\\
            correlation gain ($10 \, \mathrm{log}_{10}(B\,T_m)$) & 60 dB& 60 dB\\
             transmit Power & 10 dBm & 1 dBm\\
            antenna gain Rx & 4 dBi& 9 dBi\\
            antenna gain Tx & 4 dBi& 7 dBi\\
            VUCA radius ($R_A$)&0.144 m& 0.085 m\\
            polarization  & vertical & vertical\\
			\bottomrule
        \end{NiceTabular}
    \end{center}
\end{table}
%
\begin{figure*}[t]
  \centering
  \includegraphics[width = 2\columnwidth]{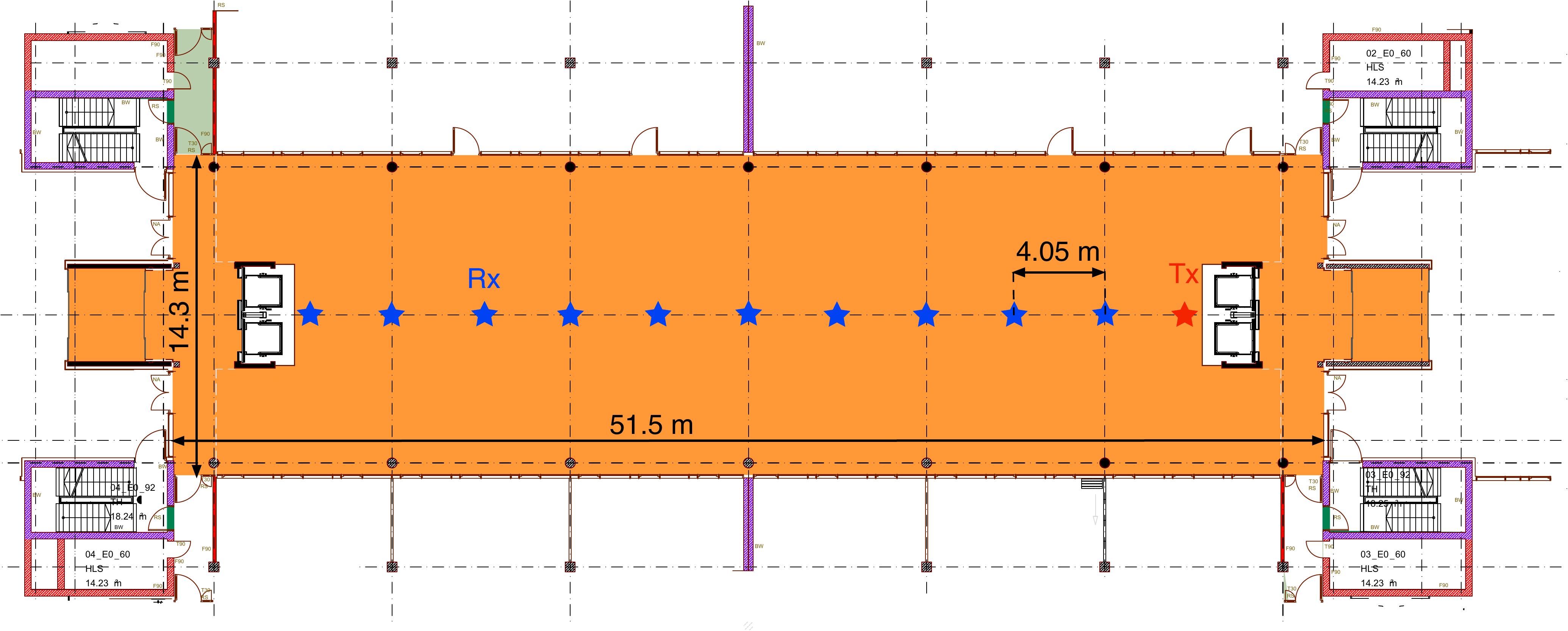}
  \caption{Plan of the measurement scenario}
  \label{fig:plan}
 \end{figure*}
 
\begin{figure}[htbp]
  \centering
  \includegraphics[width = 0.9\columnwidth]{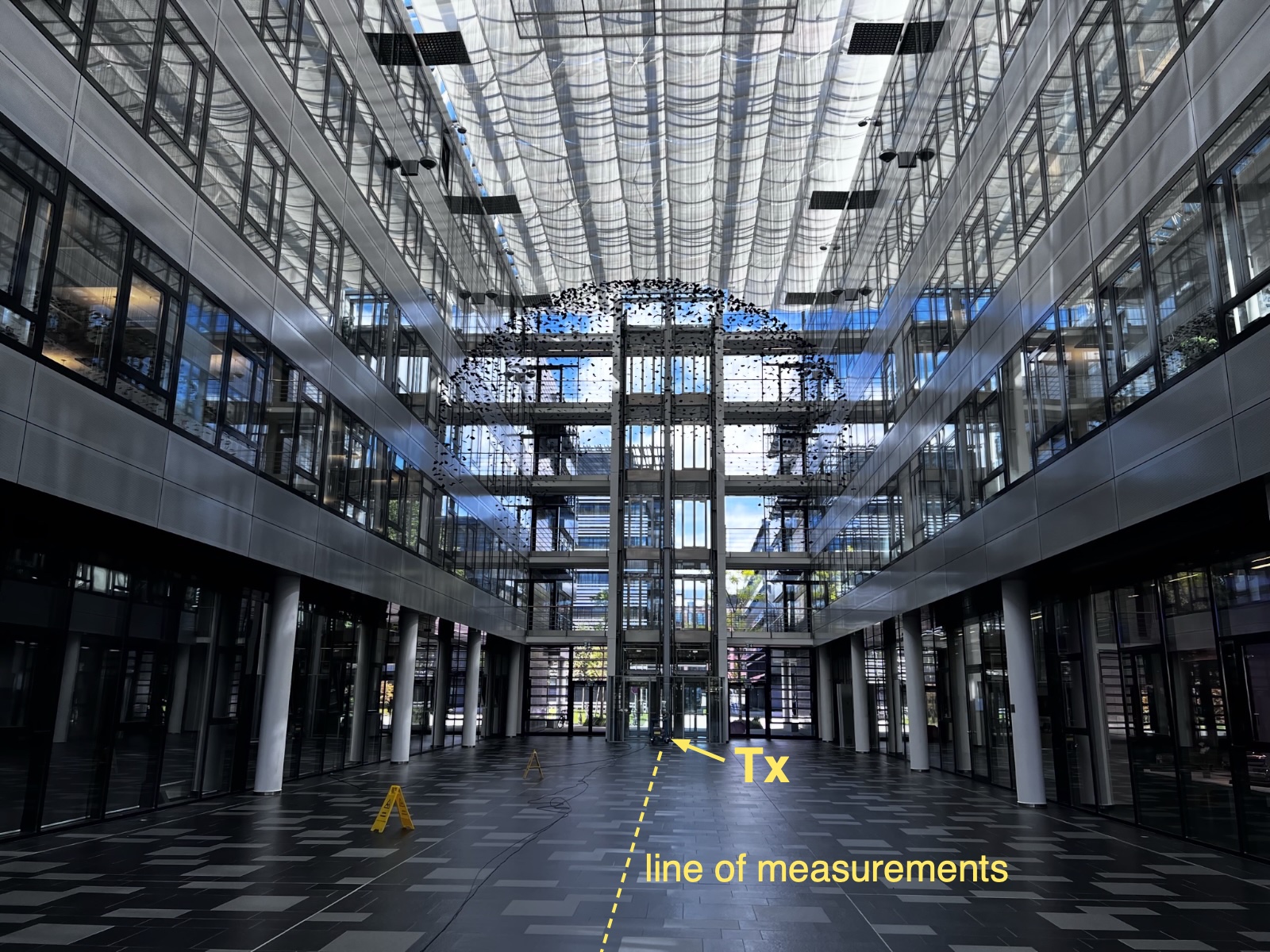}
  \caption{Indoor measurement scenario, indicating the Tx position and the line of Rx positions}
  \label{fig:scenario}
\end{figure}

The measurements were performed in a large open space or ``atrium'' of an office building, shown in Fig. \ref{fig:scenario}, which is similar to a shopping mall. The atrium has metal panels and glass windows on each side and glass walls and elevators in front and in the back as well as a glass roof. The overall area is 740 $\mathrm{m}^2$, the height is 22 m,  the distance between the elevators is 41 m, and the width of the atrium is 14.3 m. For the measurements the Tx was fixed near one elevator and the Rx was moved along a straight line (see Fig. \ref{fig:plan}), capturing 10 test points (TPs) at both frequencies, with an average spacing of 4.05 m, starting at a minimum Tx distance of 3.30 m (TP 1) and going up to 39.75 m (TP 10).
%

\section{Experimental Results}
The captured IQ-Data were read from the signal analyzer and post-processed in MatLab\textregistered. The first processing steps included the application of the back-to-back calibration measurement, the correlation and filtering in frequency domain (using a ultra-spherical filter with power sidelobe level (PSL) and slope taper rate $\alpha$) as well as oversampling in delay domain, as described in \cite{keusgen}. Additionally, a low-pass filter (Tuckey filter with length $L_D$ and parameter $\alpha$) in the ``Doppler'' domain  -- which corresponds to the spectral domain of the array responses -- was applied. This filter removes noise and thus realizes an additional gain from spatial oversampling which adds up to the correlation gain and oversampling gain. These processing steps yield the input delay spread function (IDSF), ready for DoA estimation. Table \ref{tab:ev_parameters} records the chosen evaluation parameters.
\begin{table}[htbp]
\caption{Evaluation Parameters}
\label{tab:ev_parameters}
    \begin{center}
        \begin{NiceTabular}[c]{l l l}
        \CodeBefore
        \rowcolor{gray!50}{1-1}
       \rowcolors{2}{gray!15}{}
        \Body
            \toprule
			 \textit{Parameter}& 14 GHz & 160 GHz\\
			\midrule
   oversampl. in delay domain & 4 & 4 \\
   freq. domain filter & ultra-spheric. & ultra-spheric.\\
   freq. filter param.:  PSL, $\alpha$ & 70 dB, 1 & 70 dB, 1\\
   min. number of virt. ant. ($K_\mathrm{min})$ & 85 & 571\\
spectral domain filter & Tuckey & Tuckey\\
   spectral filter param.: $L_D$, $\alpha$ & 85, 0.5 & 571, 0.5\\
spectr. filter gain ($10 \, \mathrm{log}_{10}\left(\sfrac{K}{K_\mathrm{min}}\right)$) & 10.7 dB & 3.9 dB\\
  realized noise floor &$\sim $ -126 dB& $\sim $ -117 dB\\
   relative threshold & 25 dB & 25 dB\\
delay clustering & 0.25 ns & 0.25 ns\\
angular clustering& 6°& 6°\\
resolution BS-MUSIC&0.05°&0.05° \\
			\bottomrule
        \end{NiceTabular}
    \end{center}
\end{table}
\subsection{Illustration of DoA Estimation}
\begin{figure}[htbp]
    \setlength{\abovecaptionskip}{0pt plus 0pt minus 0pt}
    \setlength{\belowcaptionskip}{0pt plus 0pt minus 0pt}
    \vspace{-0mm}
    \centering
    \includegraphics[scale = 0.9]{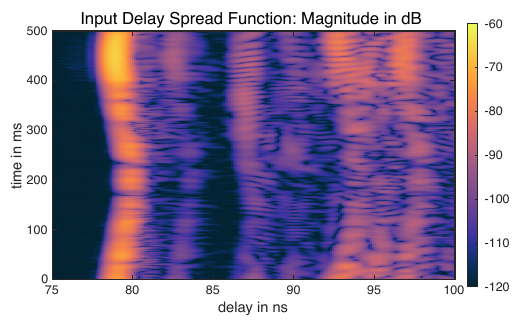}
    \caption{Detail of magnitude of IDSF (TP 6, 14 GHz)}
    \label{fig:idsf}
    \vspace{-0mm}
\end{figure}

\begin{figure}[htbp]
\setlength{\abovecaptionskip}{0pt plus 0pt minus 0pt}
    \setlength{\belowcaptionskip}{0pt plus 0pt minus 0pt}
\vspace{-0mm}
  \centering
 \includegraphics[scale = 0.9]{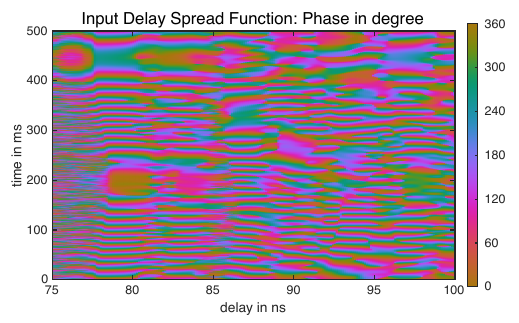}
  \caption{detail of phase of IDSF (TP 6, 14 GHz)}
  \label{fig:idsf_phase}
  \vspace{-0mm}
\end{figure}

Figures \ref{fig:idsf} and \ref{fig:idsf_phase} show a detail of magnitude and phase of a typical IDSF for 14 GHz after performing all calibration and filtering (including oversampling).
One can clearly recognize the LOS path at around 79 ns and its change in delay due to the displacement of the antenna and also the induced ``Doppler shift'', due to the movement. One can at least guess the (unwraped) sinusoidal phase of the strongest peak from the phase plot. 

\begin{figure}[htbp]
\setlength{\abovecaptionskip}{0pt plus 0pt minus 0pt}
    \setlength{\belowcaptionskip}{0pt plus 0pt minus 0pt}
\vspace{-0mm}
  \centering
  \includegraphics[scale = 0.9]{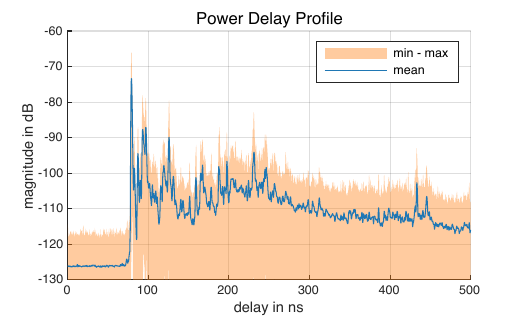}
  \caption{Power Delay Profile (TP 6, 14 GHz)}
  \label{fig:pdp}
  \vspace{-0mm}
\end{figure}

\begin{figure}[htbp]
\setlength{\abovecaptionskip}{0pt plus 0pt minus 0pt}
    \setlength{\belowcaptionskip}{0pt plus 0pt minus 0pt}
\vspace{-0mm}
  \centering
  \includegraphics[scale = 0.9]{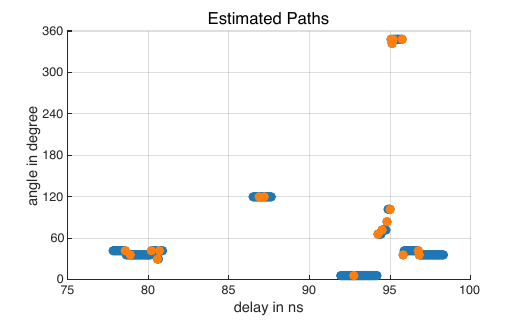}
  \caption{Clustering (detail) in the delay-azimuth plane (TP 6, 14 GHz)}
  \label{fig:clustering}
  \vspace{-0mm}
\end{figure}

\begin{figure}[htbp]
\setlength{\abovecaptionskip}{0pt plus 0pt minus 0pt}
    \setlength{\belowcaptionskip}{0pt plus 0pt minus 0pt}
\vspace{-0mm}
  \centering
  \includegraphics[scale = 0.9]{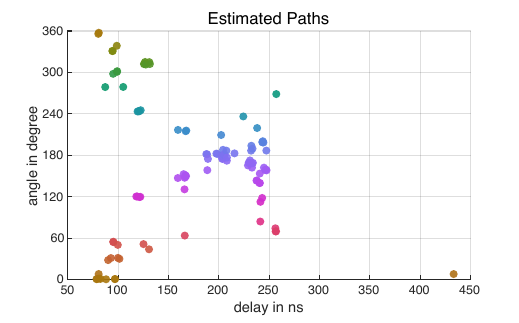}
  \caption{Estimated paths in delay-azimuth plane (TP 6, 14 GHz)}
  \label{fig:az_delay}
  \vspace{-0mm}
\end{figure}

\begin{figure*}
  \begin{center}
    \begin{subfigure}[t]{8.8cm}
    \centering
      \includegraphics[scale = 0.66]{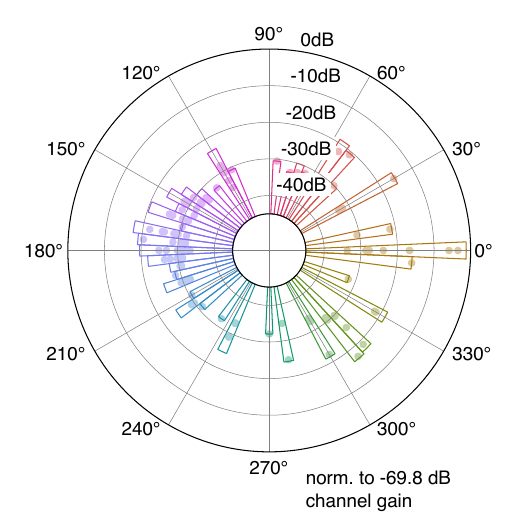}
  \caption{14 GHz}
      \label{fig:rose_14}
    \end{subfigure}
    \vline\hfill
    \begin{subfigure}[t]{8.8cm}
   \centering
    \includegraphics[scale = 0.66]{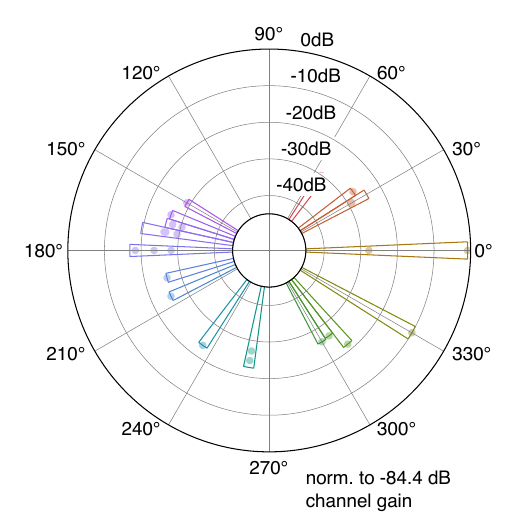}
  \caption{160 GHz}
     \label{fig:rose_160}
    \end{subfigure}
  \end{center}
  \caption{Rose plots of estimated paths of TP 6 for both frequencies}
  \label{fig:spec}
 \end{figure*}

As expected, the magnitude for each path varies, due to the non-omnidirectional antennas, which could also be seen in the power delay profile (PDP) \cite{keusgen}, in Fig. \ref{fig:pdp}. In this figure, the colored area defines the envelope of the all channel impulse responses (CIRs, i.e. the rows of the IDSF), whereas the maximum values are associated with the perfect alignment of the directive antenna towards the main direction for the respective delay sample. After applying a relative threshold with respect to the maximum envelope, the BS-MUSIC algorithm is applied to each delay sample above the threshold, estimating one main direction in azimuth and elevation, whereas just the azimuth information is processed further. Fig. \ref{fig:clustering} shows all detected paths in the delay azimuth plane, which are clustered to the intended delay and azimuth resolution (blue circles) and the estimated ``true'' paths as its local power maxima and shown as orange circles, which concludes the estimation process. The final result for this TP is shown in Fig. \ref{fig:az_delay}, Fig. \ref{fig:pdp}, and as rose plot (azimuth angle and power) in Fig. \ref{fig:rose_14}. Fig. \ref{fig:rose_160} shows the result of the same TP at 160 GHz, which indicates the excellent agreement of main directions between both frequencies, but also the lower density of paths for the higher frequency.
\begin{figure}[htbp]
\setlength{\abovecaptionskip}{0pt plus 0pt minus 0pt}
    \setlength{\belowcaptionskip}{0pt plus 0pt minus 0pt}
\vspace{-0mm}
 \centering
  \includegraphics[scale = 1]{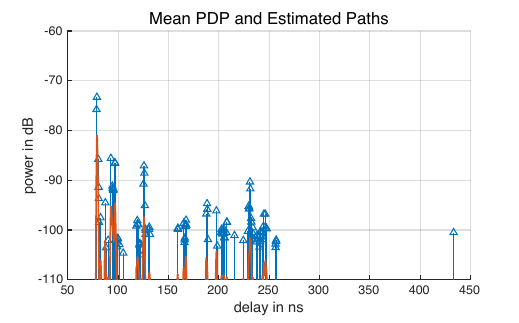}
  \caption{Estimated paths in delay domain (TP 6, 14 GHz)}
  \label{fig:example_path_pdp}
  \vspace{-0mm}
 \end{figure}

\subsection{Estimated Parameters}
\begin{figure}[htbp]
\setlength{\abovecaptionskip}{0pt plus 0pt minus 0pt}
    \setlength{\belowcaptionskip}{0pt plus 0pt minus 0pt}
\vspace{-0mm}
  \centering
  \includegraphics[scale = 1]{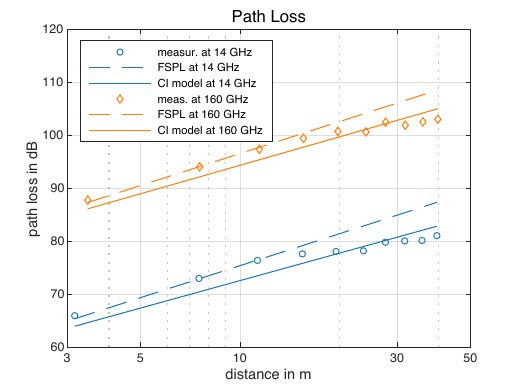}
  \caption{Estimated and modeled path loss for both frequencies}
  \label{fig:pathloss}
  \vspace{-0mm}
\end{figure}

\begin{figure}[htbp]
\setlength{\abovecaptionskip}{0pt plus 0pt minus 0pt}
    \setlength{\belowcaptionskip}{0pt plus 0pt minus 0pt}
\vspace{-0mm}
  \centering
  \includegraphics[scale = 1]{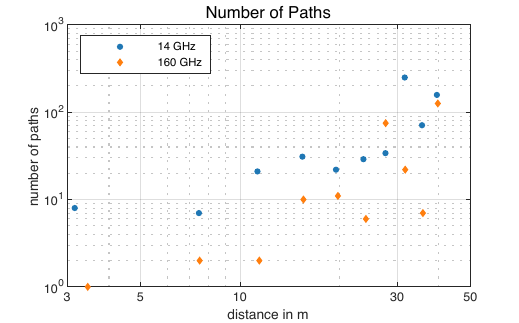}
  \caption{Estimated number of paths for both frequencies}
  \label{fig:number_paths}
  \vspace{-0mm}
\end{figure}

\begin{figure}[htbp]
\setlength{\abovecaptionskip}{0pt plus 0pt minus 0pt}
    \setlength{\belowcaptionskip}{0pt plus 0pt minus 0pt}
\vspace{-0mm}
  \centering
  \includegraphics[scale = 1]{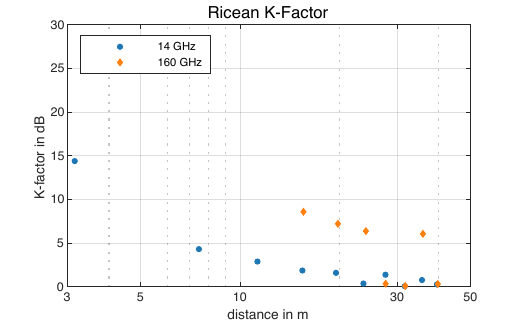}
  \caption{Estimated K-factor for both frequencies, values for first 3 TPs at 160 GHz are above 30 dB }
  \label{fig:k_factor}
  \vspace{-0mm}
\end{figure}

\begin{figure}[htbp]
\setlength{\abovecaptionskip}{0pt plus 0pt minus 0pt}
    \setlength{\belowcaptionskip}{0pt plus 0pt minus 0pt}
\vspace{-0mm}
  \centering
  \includegraphics[scale = 1]{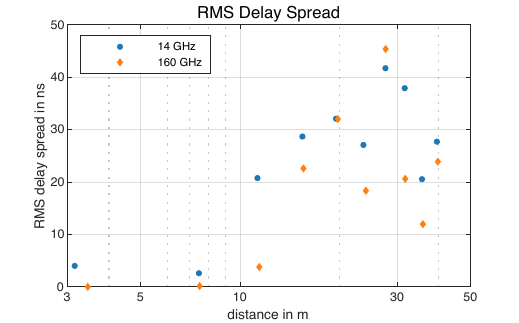}
  \caption{Estimated RMS DS for both frequencies}
  \label{fig:rms_delay}
  \vspace{-0mm}
\end{figure}

\begin{figure}[htbp]
\setlength{\abovecaptionskip}{0pt plus 0pt minus 0pt}
    \setlength{\belowcaptionskip}{0pt plus 0pt minus 0pt}
\vspace{-0mm}
  \centering
  \includegraphics[scale = 1]{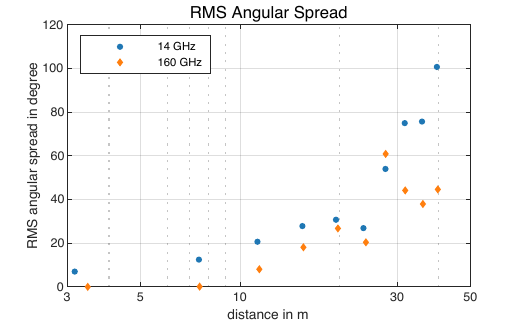}
  \caption{Estimated RMS AS for both frequencies}
  \label{fig:rms_angular}
  \vspace{-0mm}
\end{figure}

The following figures show the estimated channel parameters over distance defined in sec. \ref{sec:parameter} for all TPs and both frequencies: The PL in Fig. \ref{fig:pathloss} clearly deviates from the FSPL for larger distances due to a strong increase in the number of paths, as it can be clearly seen from Fig. \ref{fig:number_paths}, and  which is reflected in a path loss exponent $n$ of the CI model smaller than 2. As expected, the number of paths is lower for 160 GHz compared to 14 GHz. Fig. \ref{fig:k_factor} shows the K-factor, which is negatively correlated with the number of paths and thus decreases with distance. Finally, Figures \ref{fig:rms_delay} and \ref{fig:rms_angular} depict RMS DS and RMS AS which also increase strongly according to the number of paths, with a tendency to be smaller for the higher frequency. The Table \ref{tab:channel_param} summarizes the characteristic values of the estimated data.
%
%
%
%
%
%
\begin{table}[htbp]
\caption{Estimated Channel Parameters}
\label{tab:channel_param}
    \begin{center}
        \begin{NiceTabular}[c]{l | r r r | r r r}
        \CodeBefore
        \rowcolor{gray!50}{1-2} 
       \rowcolors{3}{}{gray!15}{} 
        \Body
            \toprule
			 &\multicolumn{3}{c}{14 GHz}&\multicolumn{3}{c}{160 GHz}\\
   \textit{Parameter} &  &\textit{min} &\textit{max}& &\textit{min}&\textit{max}\\
			\midrule
   PL exp. ($n$) w/o unit& 1.72 &--&--&1.78&--&--\\
     RMS DS in ns &--&2.61&41.74&--&0&45.39 \\
    RMS AS in degree&--&7.00&100.76&--&0&60.91 \\
  K-factor in dB &--&0.08& 14.41&--&0.10&$\infty$\\
  number of paths&--& 7 & 250 &-&1&126\\
			\bottomrule
        \end{NiceTabular}
    \end{center}
\end{table}
%
%
%
\section{Conclusion}
We have demonstrated a novel rotary measurement platform which can be easily adapted for different frequency regions from FR1 up to the THz range by utilizing various antennas or sub-THz and THz front-ends. The system has been applied to perform fast, high angular resolution directional channel measurements at 14 GHz (FR3) and at 160 GHz in an indoor atrium-type hall environment. The measurements exhibit significant multipath components even for the sub-THz frequencies and the directional polar plots show a very close correlation of both frequencies, underlining the validity of the dual-frequency measurement. 
The estimated number of paths as well as delay and angular spread increase strongly with distance for both frequencies.



\end{document}